\documentclass[aps,pra,twocolumn,superscriptaddress]{revtex4-1}

\usepackage{graphicx}
\usepackage{amsmath}
\usepackage{caption}
\usepackage{graphicx}
\usepackage{float} 
\usepackage{subfigure}
\usepackage{verbatim}
\usepackage{float}

\begin{document}

\title{Delta interferometer: A polarization sensitive interferometer}

\author{Chaoqi Wei}
\affiliation{Electronic Materials Research Laboratory, Key Laboratory of the Ministry of Education \& International Center for Dielectric Research, Xi'an Jiaotong University, Xi'an 710049, China}

\author{Jianbin Liu}
\email[]{liujianbin@xjtu.edu.cn}
\affiliation{Electronic Materials Research Laboratory, Key Laboratory of the Ministry of Education \& International Center for Dielectric Research, Xi'an Jiaotong University, Xi'an 710049, China}
\affiliation{The Key Laboratory of Weak Light Nonlinear Photonics (Nankai University, Tianjin 300457), Ministry of Education, China}

\author{Yunong Sun}
\affiliation{Electronic Materials Research Laboratory, Key Laboratory of the Ministry of Education \& International Center for Dielectric Research, Xi'an Jiaotong University, Xi'an 710049, China}

\author{Rui Zhuang}
\affiliation{Electronic Materials Research Laboratory, Key Laboratory of the Ministry of Education \& International Center for Dielectric Research, Xi'an Jiaotong University, Xi'an 710049, China}

\author{Yu Zhou}
\affiliation{MOE Key Laboratory for Nonequilibrium Synthesis and Modulation of Condensed Matter, Department of Applied Physics, Xi’an Jiaotong University, Xi’an, Shaanxi 710049, China}

\author{Huaibin Zheng}
\affiliation{Electronic Materials Research Laboratory, Key Laboratory of the Ministry of Education \& International Center for Dielectric Research, Xi'an Jiaotong University, Xi'an 710049, China}

\author{Yanyan Liu}
\affiliation{Science and Technology on Electro-Optical Information Security Control Laboratory, Tianjin, 300308, China}

\author{Xiusheng Yan}
\affiliation{Science and Technology on Electro-Optical Information Security Control Laboratory, Tianjin, 300308, China}

\author{Fuli Li}
\affiliation{MOE Key Laboratory for Nonequilibrium Synthesis and Modulation of Condensed Matter, Department of Applied Physics, Xi’an Jiaotong University, Xi’an, Shaanxi 710049, China}

\author{Zhuo Xu}
\affiliation{Electronic Materials Research Laboratory, Key Laboratory of the Ministry of Education \& International Center for Dielectric Research, Xi'an Jiaotong University, Xi'an 710049, China}

\date{\today}

\begin{abstract}
A new type of polarization sensitive interferometer is proposed, which is named as Delta interferometer due to the geometry of the simplest interferometer looks like Greek letter, Delta. To the best of our knowledge, it is the first time that this type of interferometer is proposed. The main difference between Delta interferometer and other existed interferometer, such as Michelson interferometer, Mach-Zehnder interferometer, Young\rq{}s double-slit interferometer, is that the two interfering paths are asymmetrical in Delta interferometer, which makes it polarization sensitive. The visibility of the first-order interference pattern observed in Delta interferometer is dependent on the polarization of the incidental light. Optical coherence theory is employed to interpret the phenomenon and a single-mode continuous-wave laser is employed  to verify the theoretical predictions. The theoretical and experimental results are consistent. Delta interferometer provides a perfect tool to study the reflection of electric field in different polarizations and may find applications in polarization sensitive scenarios.
\end{abstract}

\maketitle

\section{Introduction}\label{introduction}

Interferometer plays an essential role in the development of physics, especially optics. For instance, light was first thought as consisting of particles \cite{newton-optics}. Young\rq{}s famous interfering experiments confirmed that light consists of wave instead of particles \cite{young-1804}. Later, it was proved that the first-order interference pattern exists when single photons were input into Mach-Zehnder interferometer, which indicates the wave-particle duality of photon \cite{aspect-1986}. Another example is that Michelson–Morley experiment confirmed that there is no such thing called ether by employing light in a Michelson interferometer \cite{michelson}, which eventually leads to the generation of special relativity \cite{einstein}. All the existed interferometers, such as Young\rq{}s double-slit interferometer, Michelson interferometer, Mach-Zehnder interferometer, Sagnac interferometer, Fabry-Perot interferometer and so on \cite{interferometer-book}, are different in schemes. However, all the interferometers mentioned above have one important property in common, which is that the difference of light reflections in the two interfering paths is even. This property guarantees that the polarizations of these two interfering beams are identical in the observation plane no matter what the input polarization is.  In other words, these optical interferometers are insensitive to the polarization of input light.

In classical electromagnetic theory, it is well-known that the reflection of electric field is dependent on the polarization of the input field \cite{EM-book1, EM-Yang}. What happens if the difference between light reflections of the two interfering paths is odd? Is the visibility of interference pattern varies as the polarization of the input light changes? In order to answer these questions, we propose a new type of interferometer, in which the difference of the light reflections between the two interfering paths is odd. The simplest interferometer of this type, which is called Delta interferometer due to its shape looks like the Greek letter, $\Delta$, is employed to study how the first-order interference pattern changes as the polarization of the input light varies. 

There exists a type of interferometer called polarization interferometer  \cite{pm-book, pm-rev, pm-ao, pm-cr}. The difference between polarization interferometer and the proposed Delta interferometer are as follows. In a typical polarization interferometer, two orthogonally polarized electric fields are usually  employed and the polarization difference are created by polarization components, such as polarization beam splitter \cite{pm-rev}, quarter wave-plate \cite{pm-ao}, and so on. In Delta interferometer, the polarizations of the two interfering electric fields are not always orthogonal. The difference in polarizations of interfering electric fields in Delta interferometer is created by the reflection time difference. A more important difference between these two types of interferometers lies at how the interference patterns are obtained. In polarization interferometer, two orthogonally polarized fields are usually measured after passing a polarizer oriented in $45^\circ$ \cite{pm-2000, pm-cr} or measured independently and then calculate the difference between these two measured signals \cite{pm-1987}. While in Delta interferometer, the two interfering electric fields are combined directly to observe the first-order interference pattern, which is more like Mach-Zehnder interferometer instead of polarization interferometer. Delta interferometer provides a perfect scheme to study the reflection of electric field in different polarizations and may find possible applications in polarization sensitive scenario, such as the observation of phase object \cite{born-book}, polarization modulated two-photon interference \cite{luo-2021}, and so on.

The remaining parts of the paper are organized as follows. The introduction of Delta interferometer and calculations of the first-order interference pattern in Delta interferometer are in Sect. \ref{theory}. Section \ref{experiments} includes the experiments to verify the theoretical predictions. The discussions and conclusions are in Sects. \ref{discussions} and \ref{conclusions}, respectively.

\section{Theory}\label{theory}

\begin{figure}[htb]
\centering
\includegraphics[width=55mm]{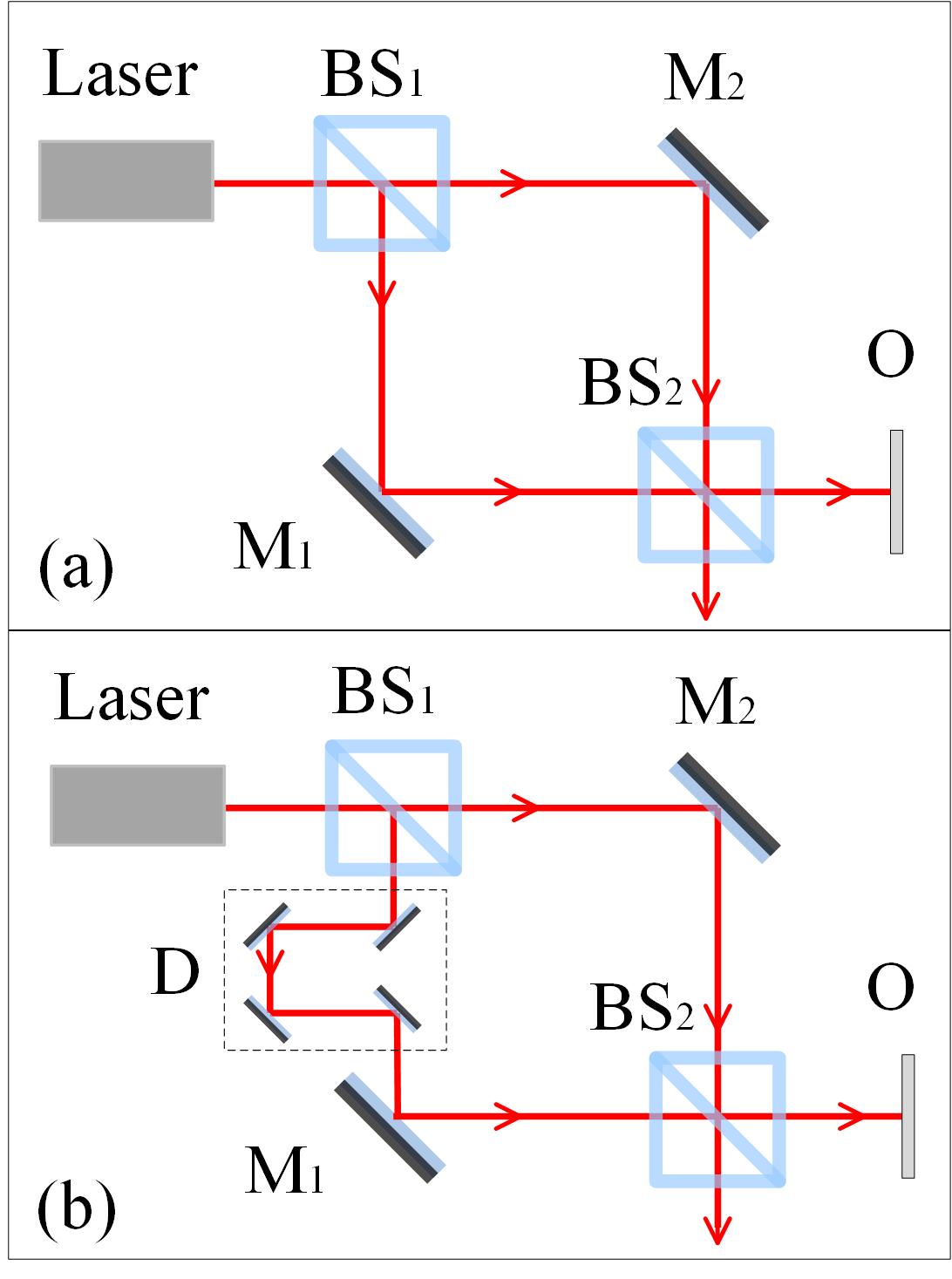}%
\caption{Mach-Zehnder interferometer (a) and modified Mach-Zehnder interferometer (b). Laser is single-mode continuous-wave laser. BS is non-polarizing beam splitter. M is mirror. O is observation plane. D represents delay line in the modified Mach-Zehnder interferometer consisting of four mirrors.}\label{1}
\end{figure}

For comparison,  Mach-Zehnder (M-Z) interferometer and modified M-Z interferometer shown in Fig. \ref{1} are taken as an example to show the difference between the existed interferometers and our proposed Delta interferometer. In the M-Z interferometer shown in Fig. \ref{1}(a), there are two interfering paths. One is the light transmitting through beam splitter 1 (BS$_1$), reflected by mirror 2 (M$_2$), and then reflected by BS$_2$. Light traveling along this path is reflected twice. Light traveling in the other path is reflected by  BS$_1$ and M$_1$, respectively. The difference between the time of reflection in these two paths equals zero, which is an even number. In the modified M-Z interferometer shown in Fig. \ref{1}(b), the difference between the time of light reflection in the two interfering paths equals 4, which is also an even number. To the best of our knowledge, the difference between the time of light reflection in two interfering paths is even for all the well-known interferometers, such as Young\rq{}s double-slit interferometer, Michelson interferometer, Mach-Zehnder interferometer, Sagnac interferometer, Fabry-Perot interferometer and so on \cite{interferometer-book}. 

\begin{figure}[htb]
\centering
\includegraphics[width=60mm]{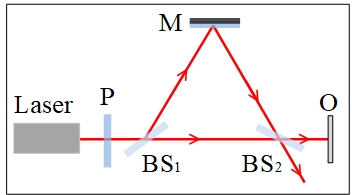}%
\caption{Delta interferometer. Laser is single-mode continuous-wave laser. P is half wavelength plate. BS is non-polarizing beam splitter. M is mirror. O is observation plane.}\label{2}
\end{figure}

Figure \ref{2} shows the scheme of the proposed Delta interferometer.  Laser is a single-mode continuous-wave laser. P is a half wavelength plate employed to control the polarization of laser light. BS is non-polarizing beam splitter, in which plate beam splitter instead of cubic beam splitter is employed in Delta interferometer. M is mirror and O is the observation plane. There are two paths for laser light reaching the observation plane in Delta interferometer. One path is light transmitting through BS$_1$ and BS$_2$. The other path is light reflected by BS$_1$,  M, and BS$_2$, respectively. The difference between the time of light reflection in the two interfering paths equals 3, which is an odd number.

\begin{figure}[htb]
\centering
\includegraphics[width=55mm]{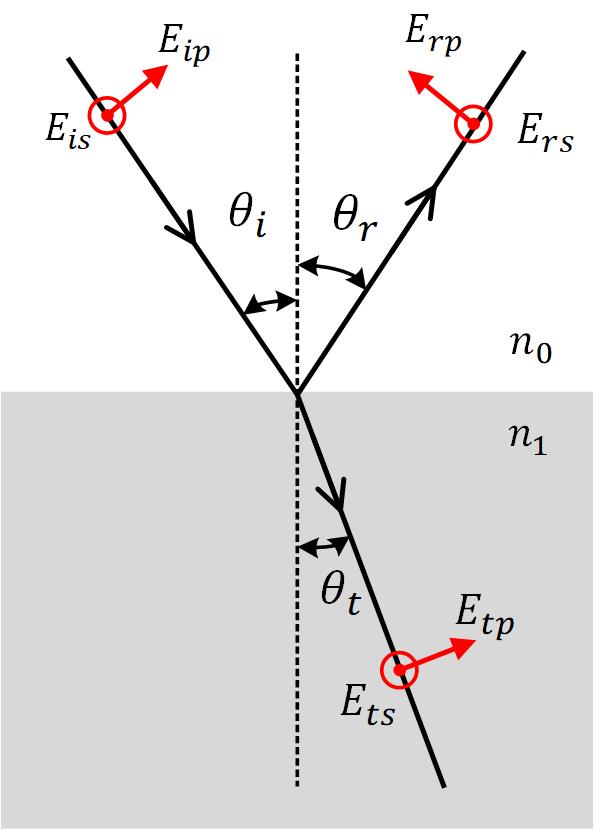}%
\caption{Refection and transmission of electric field on the boundary between two medium. $n_0$ and $n_1$ are the refractive index of these two medium. $E_{ip}$, $E_{rp}$, and $E_{tp}$ are the incidental, reflected, and transmitted electric fields parallel to the reflection plane, respectively.  $E_{is}$, $E_{rs}$, and $E_{ts}$ are the  incidental, reflected, and transmitted electric fields perpendicular to the reflection plane, respectively. $\theta_i$, $\theta_r$, and $\theta_t$ are the incidental, reflection, and transmission angles, respectively. }\label{3} 
\end{figure}

In order to study the first-order interference pattern in Delta interferometer, one has to take the reflection and transmission in different polarization states into account. We will follow the custom introduced in Ref. \cite{handbook} to discuss the reflection and transmission of light field. Figure \ref{3} shows electric field with arbitrary polarization being incident onto a boundary between a medium of refractive index $n_0$ and a medium of refractive index $n_1$. Since any polarized electric field can be divided into the superposition of two orthogonally polarized fields \cite{EM-book1}. It is convenient to choose these two orthogonally directions being parallel and perpendicular to the reflection plane, respectively. $E_{ip}$ and $E_{is}$ represents the parallel and perpendicular parts of the incident electrical field, respectively. $E_{rp}$, $E_{rs}$, $E_{tp}$, and $E_{ts}$ are defined similarly, in which $r$ and $t$ in the subscript are short for reflection and transmission, respectively. $\theta_i$, $\theta_r$, and $\theta_t$ are the incidental, reflection, and transmission angles, respectively.

The reflection and transmission coefficients of parallel and perpendicular parts are dependent on the incidental angle,  the polarization of the incidental field, and the refractive indexes of the media, which can be expressed as \cite{handbook}
\begin{eqnarray}\label{re-tr}
&&R_s = \frac{E_{rs}}{E_{is}}=\frac{n_0 \cos\theta_i-n_1\cos\theta_t}{n_0 \cos\theta_i+n_1\cos\theta_t},\nonumber\\ \label{Rs}
&&R_p= \frac{E_{rp}}{E_{ip}}=\frac{n_1 \cos\theta_i-n_0\cos\theta_t}{n_1 \cos\theta_i+n_0\cos\theta_t}, \\ 
&&T_s=\frac{E_{ts}}{E_{is}}=\frac{2n_0 \cos\theta_i}{n_0 \cos\theta_i+n_1\cos\theta_t},\nonumber\\
&&T_p=\frac{E_{tp}}{E_{ip}}= \frac{2n_0 \cos\theta_i}{n_1 \cos\theta_i+n_0\cos\theta_t}.\nonumber
\end{eqnarray}
$R_s$ and $R_p$ are the reflection coefficients of perpendicular and parallel components, respectively. $T_s$ and $T_p$ are the transmission coefficients of perpendicular and parallel components, respectively. The above equations are the well-known Fresnel formula in classical electromagnetic theory \cite{born-book, handbook}.

The electric field of linearly polarized light emitted by a single-mode continuous-wave laser can be expressed as
\begin{equation}
\vec{E}_l=\vec{E}_{pl}+e^{i\Phi}\vec{E}_{sl},
\end{equation}
where $\Phi$ is the phase difference between $\vec{E}_{pl}$ and $\vec{E}_{sl}$. $\vec{E}_{pl}$ and $\vec{E}_{sl}$ are the electric fields parallel and perpendicular to the reflection plane, respectively. 

The electric field in the observation plane are the sum of two components. One component is the field traveling through BS$_1$ and BS$_2$ (will be written as path 1 for short),
\begin{eqnarray}\label{E_o1}
&&\vec{E}_{O1}=\vec{E}_{O1p}+\vec{E}_{O1s}\\
&=&\vec{E}_{pl}(t_1)\times T_p(\text{BS}_1)\times T_p(\text{BS}_2) \nonumber\\
&&\times \exp{\{-i[\omega (t-t_1) - \vec{k}_1\cdot (\vec{r}_1-\vec{r}_0)]\}}\nonumber\\
&&+\vec{E}_{sl}(t_1)\times T_s(\text{BS}_1)\times T_s(\text{BS}_2) \nonumber\\
&&\times \exp{\{-i[\omega (t-t_1) - \vec{k}_1\cdot (\vec{r}_1-\vec{r}_0)]\}}\nonumber,
\end{eqnarray}
where $\vec{E}_{O1p}$ and $\vec{E}_{O1s}$ are the parallel and perpendicular components of  $\vec{E}_{O1}$, respectively. $T_p(\text{BS}_j)$ and $T_s(\text{BS}_j)$ are the transmission coefficients of parallel and perpendicular components for BS$_j$ ($j=1$ and 2), respectively. $\omega$ and $\vec{k}$ are the angular frequency and wave vector of the electric field, respectively. $t_1$ is the starting time of electric field in path 1. $\vec{r}_1$ is the position vector along path 1. For simplicity, the laser light is considered as single frequency in the following calculations. 

The other component in the observation plane is the field reflected by BS$_1$, M, and BS$_2$ (will be taken as path 2 for short),
\begin{eqnarray}\label{E_o2}
&&\vec{E}_{O2}=\vec{E}_{O2p}+\vec{E}_{O2s}\\
&=&\vec{E}_{pl}(t_2)\times R_p(\text{BS}_1)\times  R_p(\text{M})\times R_p(\text{BS}_2)\nonumber\\
&& \times \exp{\{-i[\omega (t-t_2) - \vec{k}_2\cdot (\vec{r}_2-\vec{r}_0)]\}}\nonumber\\
&&+\vec{E}_{sl}(t_2)\times R_s(\text{BS}_1)\times R_s(\text{M})\times R_s(\text{BS}_2)\nonumber\\
&& \times \exp{\{-i[\omega (t-t_2) - \vec{k}_2\cdot (\vec{r}_2-\vec{r}_0)]\}}\nonumber,
\end{eqnarray}
where the meanings of all the symbols are similar as the ones in Eq. (\ref{E_o1}).

Since the electric field with orthogonal polarizations can not interfere with each other \cite{born-book}, the intensity of light in the observation plane equals the sum of the parallel and perpendicular parts,
\begin{equation}\label{I_o}
I_O=|\vec{E}_{O1p}+\vec{E}_{O2p}|^2+|\vec{E}_{O1s}+\vec{E}_{O2s}|^2,
\end{equation}
where  $|...|^2$ means modulus square.  Substituting Eqs. (\ref{E_o1}) and (\ref{E_o2}) into Eq. (\ref{I_o}), it is easy to have the first-order interference pattern in the observation plane (see Appendix A for detail calculations),
\begin{eqnarray}\label{I_of}
&&I_O\\
&=&|E_{pl}|^2[|T_{p1}|^2+|R_{p2}|^2+2\text{Re}(T_{p1}R_{p2}^*e^{i\varphi})B(\Delta z)\cos(2\Delta k x)]\nonumber\\
&&+|E_{sl}|^2[|T_{s1}|^2+|R_{s2}|^2-2\text{Re}(T_{s1}R_{s2}^*e^{i\varphi})B(\Delta z)\cos(2\Delta k x)],\nonumber
\end{eqnarray}
where paraxial approximation and one-dimension are assumed to simplify the results. The assumptions $R_s(M)=-1$ and $R_p(M)=1$ for reflection on silver mirror are employed in the calculation due to silver can be approximately treated as ideal conductor  \cite{EM-Yang}. $|E_{pl}|$ and $|E_{sl}|$ are the amplitudes of electric field with parallel and perpendicular polarizations, respectively. $T_{p1}$ is short for $T_p(\text{BS}_1)\times T_p(\text{BS}_2)$. $R_{p2}$ is short for $R_p(\text{BS}_1)\times  R_p(\text{BS}_2)$. $T_{s1}$ is short for $T_s(\text{BS}_1)\times T_s(\text{BS}_2)$. $R_{s2}$ is short for $ R_s(\text{BS}_1)\times R_s(\text{BS}_2)$. $\varphi$ is the phase difference between electric field at $t_1$ and $t_2$. $B(\Delta z)$ is short for $\cos[\omega(t_1-t_2)] \cos[k_0(d_1-d_2)]$ and $\Delta z$ is proportional to the light path difference between the two interfering beams. $2\Delta k$ is the difference between $\vec{k}_1$ and $\vec{k}_2$ in the transverse direction.

Equation (\ref{I_of}) indicates that the first-order interference pattern of polarized laser light in Delta interferometer equals the sum of two sets of interference patterns. The periods of these two interference patterns are the same. However, there is a $\pi$ phase shift between these two sets of interference patterns. The position of the maximum of the first interference pattern coincides with the position of the minimum of the second interference pattern. If $|E_{pl}|^22\text{Re}(T_{p1}R_{p2}^*e^{i\varphi})$ equals $|E_{sl}|^22\text{Re}(T_{s1}R_{s2}^*e^{i\varphi})$, there will be no interference pattern observed even if there exists first-order interference pattern for parallel or perpendicular polarization individually.

\section{Experiments}\label{experiments}

In Sect. \ref{theory}, we have theoretically calculated the first-order interference pattern of polarized single-mode continuous-wave laser light in Delta interferometer. It is found that the visibility of the first-order interference pattern is dependent on the polarization of the incidental laser light. In this section, we will experimentally verify the prediction with linearly polarized laser light. 

\begin{figure}[htb]
\centering
\includegraphics[width=82mm]{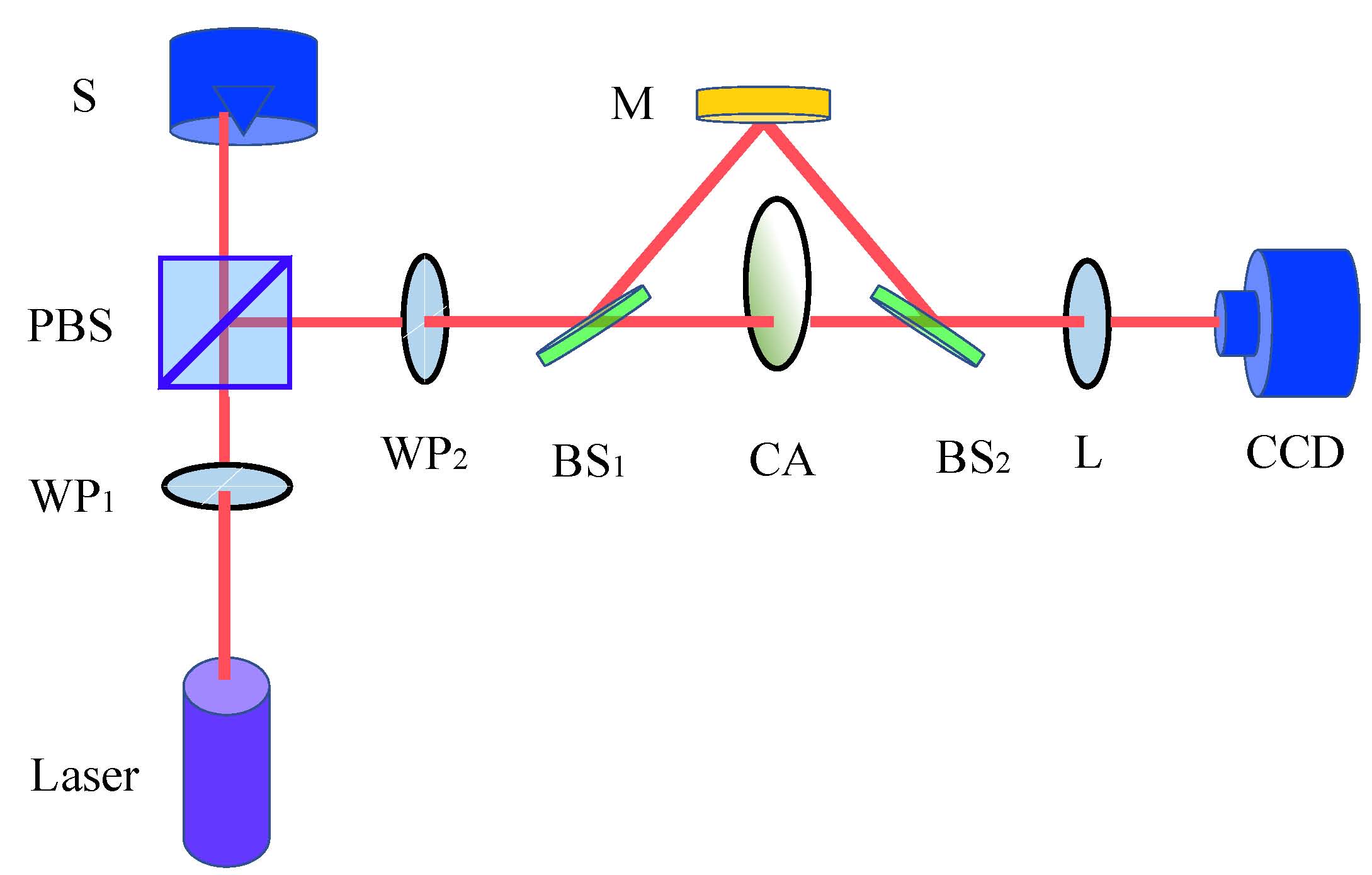}%
\caption{Experimental setup of Delta interferometer. Laser: single-mode continuous-wave laser. WP: half wavelength plate. PBS: polarizing beam splitter. S: beam stopper. BS: non-polarizing beam splitter. M: silver mirror. CA: circular intensity attenuator. L: Lens. CCD: charge-coupled device. }\label{4}
\end{figure}

The experimental setup is shown in Fig. \ref{4}. A semiconductor laser with central wavelength of 780 nm and frequency bandwidth of 200 kHz is employed as the light source. Linearly polarized light is emitted by the employed laser. A half wave-plate (WP$_1$) combined with polarizing beam splitter (PBS) are employed to control the intensity of laser light incidental into Delta interferometer.  S is a beam stopper to collect the light transmitting through PBS. The polarization of the electric field reflected by PBS is perpendicular to the reflection plane. Another half wave-plate (WP$_2$) is used to control the polarization of the incidental light. BS$_1$ and BS$_2$ are two non-polarizing 1:1 beam splitters. M is a silver mirror. CA is a circular intensity attenuator to control the intensity of transmitted light in Delta interferometer. A lens (L)  and a CCD camera are employed to record the first-order interference pattern.

\begin{figure}[htp]
	\centering
	\includegraphics[width=50mm]{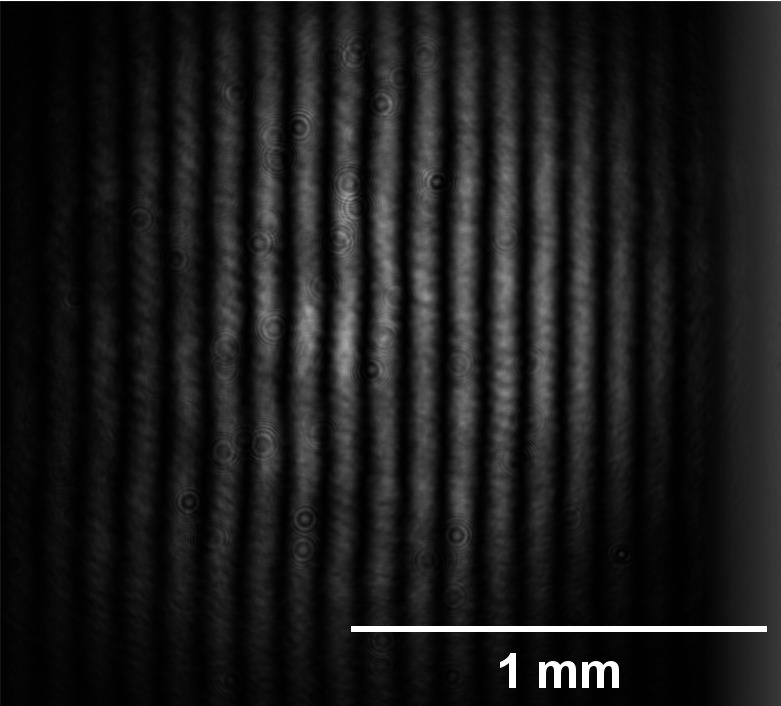}%
	\caption{Typical first-order spatial interference pattern observed in Delta interferometer.}\label{5}
\end{figure}

As mentioned in Sect. \ref{theory}, there are two interfering paths in Delta interferometer. One is reflected by BS$_1$, M, and BS$_2$. The other one is transmitted by BS$_1$ and BS$_2$.  The observed interference pattern is a result of superposition of these two light beams. A typical interference pattern in Delta interferometer is shown in Fig. \ref{5}. The observed interference pattern is a typical two-beam interference pattern, which is similar as the one observed in Michelson or M-Z interferometer when the angle between the two interfering beams are not zero.

\begin{figure}[htp]
	\centering
	\includegraphics[width=75mm]{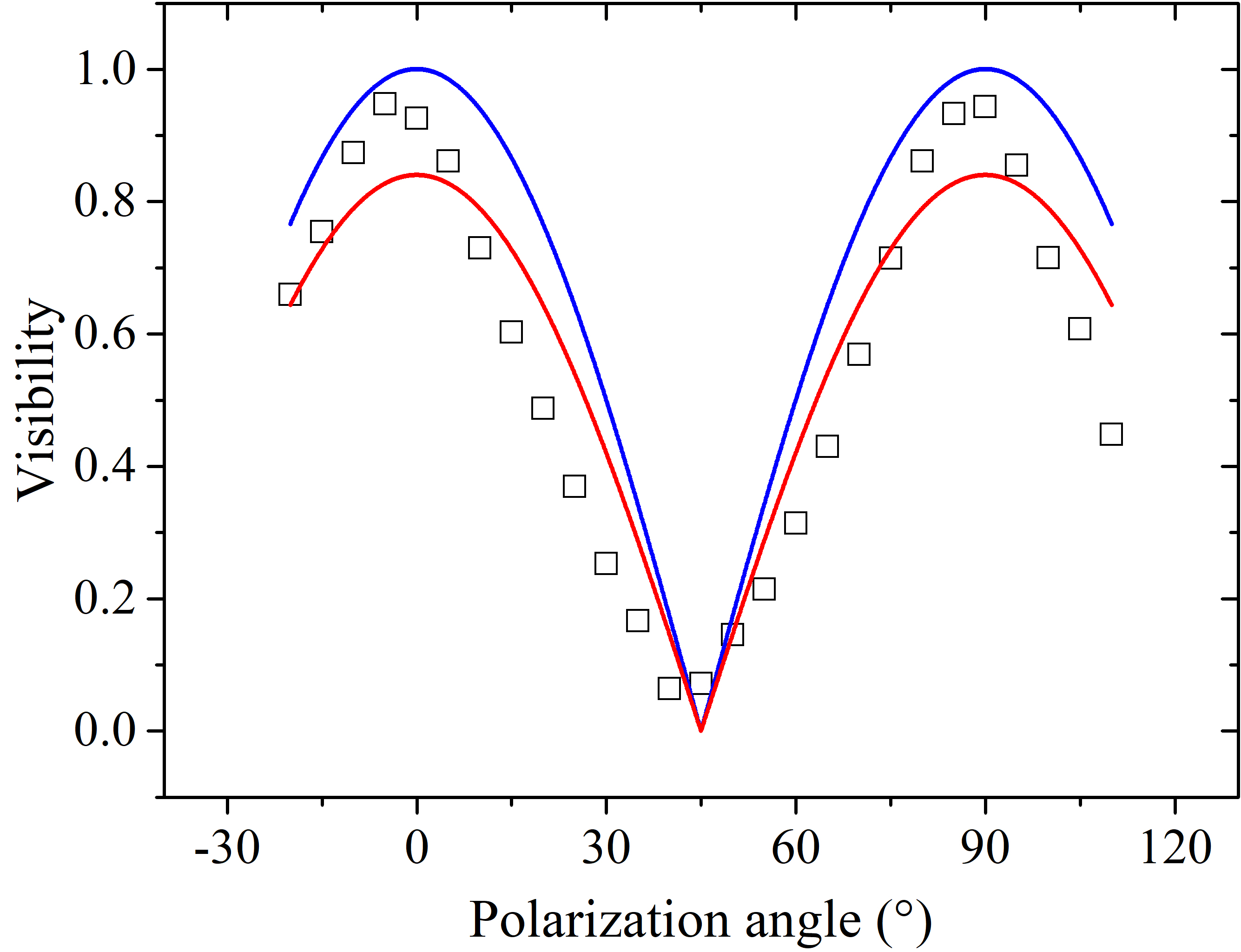}%
	\caption{(Color online) Visibility of the interference pattern with equal intensities vs. polarization angle of the input laser light. The empty squares are experimental results. The blue curve is theoretical prediction of Eq. (\ref{vis}). The red curve is theoretical fitting of the experimental data by tuning the visibility.}\label{6}
\end{figure}

Two groups interference patterns were measured. The first group interference patterns were measured by keeping the intensity of the reflected light equals the one of the transmitted light by tuning the circular intensity attenuator. In this case, the intensities of the reflected and transmitted laser light beams are made to be equal in the observation plane. The calculated visibilities of the observed interference patterns are shown in Fig. \ref{6}. The empty squares are the calculated visibilities from the observed interference patterns. The blue curve is the theoretical visibility of the interference pattern based on Eq. (\ref{I_of}), which is calculated by assuming $T_{p1}=R_{p2}$, $T_{s1}=R_{s2}$, $\varphi =0$, and $B(\Delta z)=1$. With the simplification above, it is straightforward to obtain the relation between the visibility of the interference pattern and the polarization angle of the input laser light  (see Appendix B), 
\begin{equation}\label{vis}
V_{EC}=\frac{|1-\tan ^2 \phi |}{1+ \tan ^2 \phi},
\end{equation}
where $V_{EC}$ is the calculated visibility of the interference pattern with equal intensities. $\phi$ is polarization angle of the input laser light. $\phi$ is set to be 0 for $p$ polarized light and in the range of $(-\pi/2,\pi/2]$. The red curve is the theoretical fitting of the experimental data based on Eq. (\ref{vis}) by adding a parameter, $\alpha$, in front of  $\tan ^2 \phi$ in Eq. (\ref{vis}), where $\alpha$ is in the range of $[0,1]$ and is the visibility of the interference pattern of $s$ or $p$ polarized laser light in Delta interferometer.

\begin{figure}[htp]
	\centering
	\includegraphics[width=75mm]{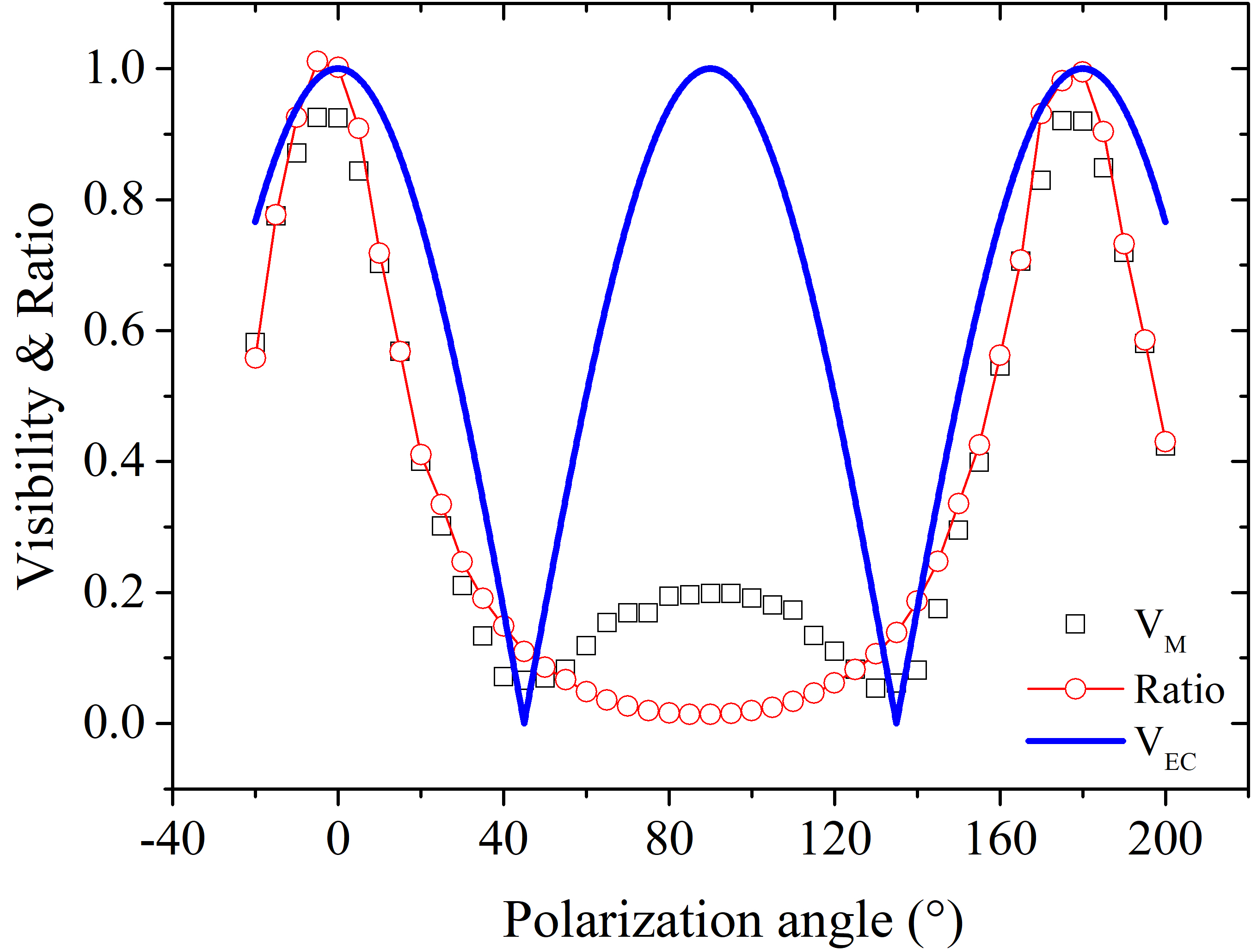}%
	\caption{(Color online) Visibility of the interference pattern with unequal intensities vs. polarization angle of the input laser light. The empty squares are measured visibility of the interference pattern. The empty circles are measured ratios between the intensities of the transmitted and reflected light beams. The blue curve is theoretical prediction of Eq. (\ref{vis}). Please see text for details.}\label{7}
\end{figure}

The second group interference patterns were measured by keeping the position of the circular intensity attenuator fixed so that the intensities of the two superposed laser light beams are equal when $p$ polarized light is incidental into the interferometer. Due to the transmission and reflection coefficients of beam splitter are dependent on the polarization angle of the incidental light, the ratio between the intensities of the two interfering light beams will change as the polarization angle of the incidental laser light changes. The black empty squares in Fig. \ref{7} are the calculated visibilities of the measured interference patten in this case. The red empty circles are the measured ratios between the intensities of the transmitted and reflected laser light beams. The red line connecting the red empty circles is served as the guidance of the eye. The blue curve is the theoretical visibility of the interference patten with equal intensities based on Eq. (\ref{vis}).

The visibility of the interference pattern in Fig. \ref{7} is dependent on the ratio of the superposed laser light beams are the visibility of the interference patten with equal intensities. When the polarization angle equals $0^\circ$, \textit{i.e.}, the input laser light is $p$ polarized, the observed visibility of the interference pattern gets its maximum value. As the polarization angle increases from $0^\circ$ to $90^\circ$, the ratio between the reflected and transmitted laser light decreases monotonously. However, the visibility of the interference pattern decreases from $0^\circ$ to $45^\circ$ and then increases from $45^\circ$ to $90^\circ$. The reason is that when the polarization angle of the input laser light equals $45^\circ$, the visibility of the interference pattern is predicted to be zero as shown by the blue curve in Fig. \ref{7}. The observed visibility in this condition is indeed the minimum, 0.066. As the polarization angle of the input laser light increases from  $45^\circ$ to $90^\circ$, the ratio between the intensities of the  transmitted and reflected laser light beams continues to decrease, while the visibility of the interference pattern increases. When the polarization angle of the input laser light equals $90^\circ$, the ratio equals its minimum, 0.014, and the visibility in this condition equals 0.20. As the polarization angle increases from $90^\circ$ to $180^\circ$, the ratio increases and the visibility gets its minimum when the polarization angle equals $135^\circ$, which is similar as the one in the $45^\circ$ case.

\section{Discussions}\label{discussions}

In last section, we have reported the visibility of the interference pattern in Delta interferometer changes as the polarization angle of linearly polarized laser light changes. A simplified theoretical result based on Eq. (\ref{vis}) is employed to interpret the observed results. Even through the tendency of the experimental results and theoretical predictions are consistent, there are some departures between  them. In this section, we will discuss why the departures exist and how to decrease the departures.

The first reason why the experimental results are different from the theoretical predictions is that the employed theoretical simplifications are usually not met in the experiments. For instance, $T_{p1}=R_{p2}$ means the overall transmission coefficient of path 1 equals the overall refection coefficient of path 2 for $p$ polarized light. $T_{s1}=R_{s2}$ means that similar conclusion holds for $s$ polarized light. These two assumptions hold only when the splitting ratio of the two employed beam splitters is $1:1$ for both $p$ and $s$ polarized light, and no energy loss for transmission and reflection. These conditions can not be met in real experiments.

The second reason is that the employed non-polarizing beam splitter (Daheng, GCC411112) is made for  $45^\circ$ incidence, which is different from the incidental angle of BS in Delta interferometer. Hence the ratio between the intensities of the transmitted and reflected light may be different from the given parameter of the beam splitter. Further more, the ratio of the beam splitter equals $1:1$ is dependent on the polarization angle of the incidental laser light, which is confirmed by the measured ratio between the intensities of the transmitted and reflected laser light shown in Fig. \ref{7}. Our theoretical model is based on the splitting ratio of beam splitter equals $1:1$  and is independent of the polarization angle of the incidental laser light, which is obviously different from the experimental conditions.

The third reason is that the transverse shape of the employed laser light beam is ellipse due to semiconductor laser is employed. Even though the total intensities of the two superposed laser light beams are equal, the intensities of the superposed two light beams may not be equal in every superposed point. It is the reason why the visibility of the observed interference pattern can not reach 1 for $s$ or $p$ polarized laser light with equal intensities shown in Fig. \ref{6}. The maximum visibility observed in our experiments equals 0.95.

The fourth reason is that silver mirror does not add an exact $\pi$ phase shift for the reflection of $s$ polarized light due to silver is not ideal conductor. When a $45^\circ$ linearly polarized laser light beam is incidental into Delta interferometer, the polarization of the reflected laser light is not $135^\circ$ linearly polarized, but elliptically polarized. There is a $45^\circ$ polarized component in the reflected light beam. The visibility of the interference pattern can not reach 0 for a $45^\circ$ linearly polarized laser light beam in Delta interferometer. In fact, the observed visibility equals 0.07 for $45^\circ$ linearly polarized laser light in Delta interferometer as shown in Fig. \ref{6}.

\begin{figure}[htp]
	\centering
	\includegraphics[width=70mm]{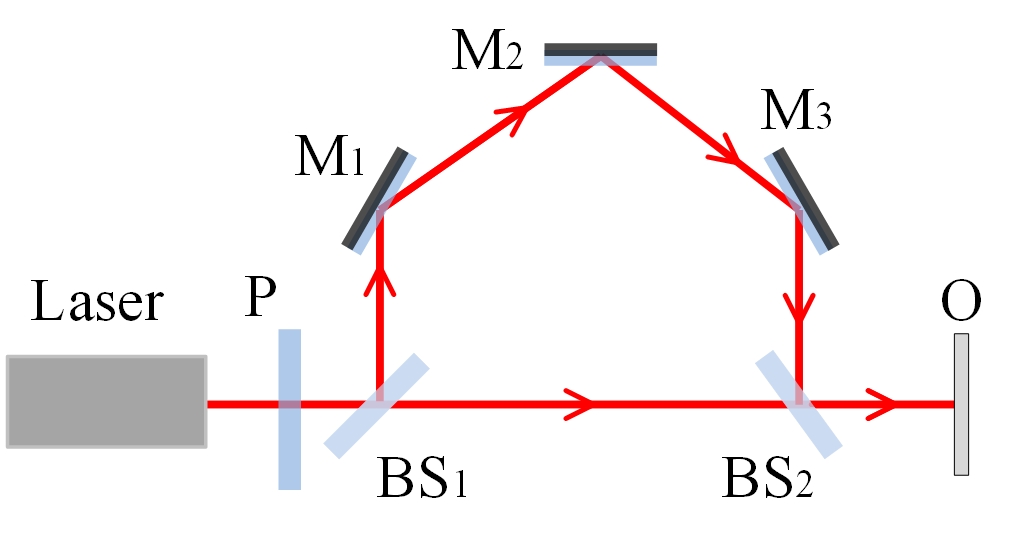}%
	\caption{Modified Delta interferometer. Comparing with Delta interferometer shown in Fig. \ref{2}, there are three mirrors instead of one. The main advantage of the modified Delta interferometer is that the incidental angle of light on BS equals $45^\circ$. All the symbols have the same meanings as the ones in Fig. \ref{2}.}\label{8}
\end{figure}

Based on the analysis above, it is straightforward to have the suggestions to optimize Delta interferometer. For instance, the main limitation listed above is the beam splitter. A modified Delta interferometer shown in Fig. \ref{8} can be employed to ensure that the incidental angle of laser light on beam splitter equals $45^\circ$ while keeping the difference between light reflections of the two interfering paths odd. As to the transverse shape of laser light, He-Ne laser instead of semiconductor laser can be employed as the light source on condition that light path difference between the two interfering beams is within the temporal coherence length of He-Ne laser. Or one change the transverse shape of semiconductor laser light beam into Gaussian with the help of different optical elements. As to the splitting ratio of beam splitter is dependent on the polarization angle of the input laser, one may find a type of beam splitter whose splitting ratio is independent of the polarization angle. Once all the conditions above were met, the departures from the theoretical and experimental results can be decreased.

\section{Conclusions}\label{conclusions}

In conclusion,we have proposed a new type of  polarization sensitive interferometer, in which the difference of light reflections between two interfering beams is odd. As an example, Delta interferometer is introduced to study the relationship between visibility of the interference pattern and the polarization angle of the input laser light. It is found that visibility of the interference pattern is indeed dependent on the polarization of the incidental laser light, which is different from Michelson or Mach-Zehnder interferometer. Two groups of interference patterns were measured and a simplified theoretical result is employed to interpret the observed the results. The experimental results and theoretical predictions are consistent. However, there are some departures. The reasons why the experimental results are different from the theoretical predictions are analyzed and corresponding suggestions to decrease the difference are given. With further improvements, this new type of polarization sensitive interferometer may serve as a basic optical interferometer in physics, especially for the demonstration of Fresnel formula. This polarization sensitive interferometer may also find potential applications in polarization sensitive application scenario, such as experiments involving birefringent crystal and so on.

\section*{Acknowledgement}
This project is supported by Shanxi Key Research and Development Project (Grant No. 2019ZDLGY09-08); Open fund of MOE Key Laboratory of Weak-Light Nonlinear Photonics (OS19-2); Fundamental Research Funds for the Central Universities.

\section*{Appendix A: Calculating the first-order interference pattern in the observation plane}\label{appendix-a}
\def\theequation{$A-$\arabic{equation}}
\setcounter{equation}{0}
\def\thefigure{$A-$\arabic{figure}}
\setcounter{figure}{0}
We will calculate the results for $|\vec{E}_{O1p}+\vec{E}_{O2p}|^2$ in detail and the results for $|\vec{E}_{O1s}+\vec{E}_{O2s}|^2$ can be obtained by analogy,
\begin{eqnarray} \label{I-ap}
&&I_{Op}=|\vec{E}_{O1p}+\vec{E}_{O2p}|^2\\ \nonumber
&=&|E_{pl}(t_1)\times T_p(\text{BS}_1)\times T_p(\text{BS}_2) \\ \nonumber
&&\times \exp{\{-i[\omega (t-t_1) - \vec{k}_1\cdot (\vec{r}_1-\vec{r}_0)]\}}\\ \nonumber
&&+E_{pl}(t_2)\times R_p(\text{BS}_1)\times  R_p(\text{M})\times R_p(\text{BS}_2)\\ \nonumber
&& \times \exp{\{-i[\omega (t-t_2) - \vec{k}_2\cdot (\vec{r}_2-\vec{r}_0)]\}}|^2, \nonumber
\end{eqnarray}
where $\vec{E}_{O1p}$ and $\vec{E}_{O2p}$ are the parallel components in Eqs. (\ref{E_o1}) and (\ref{E_o2}), respectively. For the reflection coefficient of electric field from a silver mirror, $R_p(M)$ equals 1 due to the silver can be approximately treated as ideal conductor. The refractive index $n$ equals $\sqrt{\epsilon_r\mu_r}$, where $\epsilon_r$ and $\mu_r$ are the relative permittivity and relative permeability of the medium, respectively. For dielectric material, $\mu_r$ approximately equals 1 and $\epsilon_r$ equals $\epsilon_r\rq{}-i\frac{\sigma}{\omega\epsilon_0}$, where $\epsilon_r\rq{}$ is the real part of  $\epsilon_r$, $\sigma$ is the conductivity, $\omega$ is the frequency of electric field, and $\epsilon_0$ is the permittivity of vacuum. Substituting the above expression for refractive index into Eqs. (\ref{re-tr}) and noticing $\sigma$ goes to infinity large, only the terms with $n_1$ matters. Hence it is easy to have  $R_s(M)=-1$ and $R_p(M)=1$ for the reflection of electric field from a silver mirror.

\begin{figure}[htb]
\centering
\includegraphics[width=55mm]{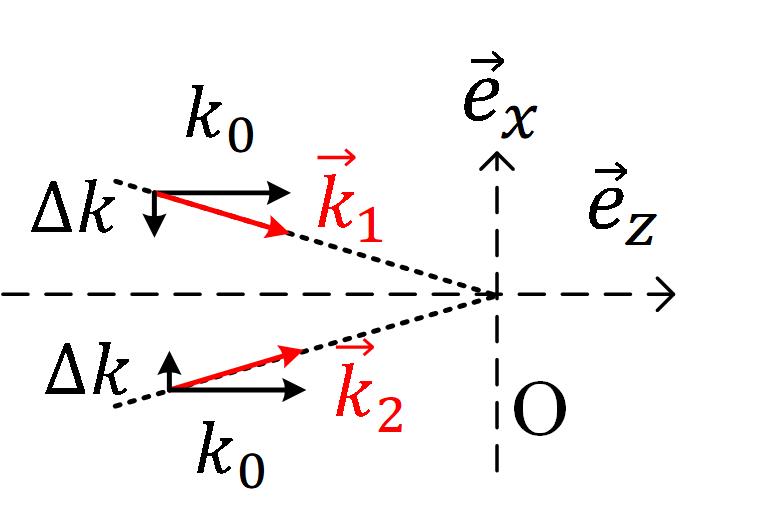}%
\caption{Symmetrical configuration for the first-order interference. $\vec{k}_1$ and $\vec{k}_2$ are the wave vectors in Eq. (\ref{a2}). $k_0$ and $\Delta k$ are the longitudinal and transverse components of the wave vector, respectively. O is the observation plane. $\vec{e}_z$ and $\vec{e}_x$ are unit vectors of longitudinal and transverse direction, respectively. One-dimension simplification is employed to calculate the transverse part.}\label{af1} 
\end{figure}

Let us wrote $T_p(\text{BS}_1)\times T_p(\text{BS}_2)$ as $T_{p1}$ and $R_p(\text{BS}_1)\times  R_p(\text{BS}_2)$ as $R_{p2}$ for short, Eq. (\ref{I-ap}) can be simplified as
\begin{eqnarray}\label{a2}
&&|E_{pl}(t_1)T_{p1}|^2+|E_{pl}(t_2)R_{p2}|^2+2\text{Re}[{E_{pl}(t_1)T_{p1}E_{pl}^*(t_2)R_{p2}^*}\nonumber \\ 
&&\times e^{-i[\omega (t_1-t_2) - (\vec{k}_1\cdot \vec{r}_1-\vec{k}_2\cdot \vec{r}_2)}],
\end{eqnarray}
where $\vec{r}_0$ is set to be 0 for simplicity and can be done by setting the origin of the coordinate system at the output position of the employed laser light. The former two terms in Eq. (\ref{a2}) are constant background in the interference pattern. The last term corresponds to the interference part, which can be further simplified by considering the paraxial approximation in the symmetrical configuration showing in Fig. \ref{af1}.

With the help from Fig. \ref{af1}, $\vec{k}_1\cdot \vec{r}_1-\vec{k}_2\cdot \vec{r}_2$ can be simplified as
\begin{eqnarray}\label{a3}
&&(k_0\vec{e}_z-\Delta k \vec{e}_x)\cdot(d_1 \vec{e}_z+x_1\vec{e}_x)\nonumber\\
&-&(k_0\vec{e}_z+\Delta k \vec{e}_x)\cdot(d_2 \vec{e}_z+x_2\vec{e}_x),
\end{eqnarray}
where $k_0$ and $\Delta k$ are the longitudinal and transverse components of the wave vector, respectively, $d_j$ and $x_j$ are the longitudinal and transverse components of the position vector $\vec{r}_j$ ($j=1$ and 2), respectively, $\vec{e}_z$ and $\vec{e}_x$ are the unit vectors in longitudinal and transverse directions, respectively, O is the observation plane. One dimension is assumed in the transverse part. Equation (\ref{a3}) can be calculated as $k_0(d_1-d_2)-\Delta k (x_1+x_2)$. For the first-order interference pattern, $x_1$ always equals $x_2$ and will be wrote as $x$ hereafter. Hence, Eq. (\ref{a3}) can be simplified as $k_0(d_1-d_2)-2\Delta k x$. 

Equation (\ref{a2}) can be simplified as
\begin{eqnarray}\label{a4}
&&|E_{pl}(t_1)T_{p1}|^2+|E_{pl}(t_2)R_{p2}|^2+2\text{Re}[{E_{pl}(t_1)T_{p1}E_{pl}^*(t_2)R_{p2}^*}]\nonumber \\ 
&&\times \cos[\omega(t_1-t_2)] \cos[k_0(d_1-d_2)] \cos(2\Delta k x),
\end{eqnarray}
where $\cos[\omega(t_1-t_2)]$ corresponds to the temporal interference, $\cos[k_0(d_1-d_2)] $ corresponds to the longitudinal spatial interference, and $\cos(2\Delta k x)$ corresponds to the transverse spatial interference. 

By analogy of the results above, $|\vec{E}_{O1s}+\vec{E}_{O2s}|^2$  can be calculated as
\begin{eqnarray}\label{a5}
&&|E_{sl}(t_1)T_{s1}|^2+|E_{sl}(t_2)R_{s2}|^2-2\text{Re}[{E_{sl}(t_1)T_{s1}E_{sl}^*(t_2)R_{s2}^*}]\nonumber \\ 
&&\times \cos[\omega(t_1-t_2)] \cos[k_0(d_1-d_2)] \cos(2\Delta k x),
\end{eqnarray}
where the plus sign between constant background and the interference term in Eq. (\ref{a4}) changes into minus sign in Eq. (\ref{a5}) due to  $R_p(M)=1$ and  $R_s(M)=-1$. 

The final interference pattern in the observation plane is the sum of these  two interference patterns with the same period. However, there is a $\pi$ phase shift between these two sets of patterns. If $E_{pl}(t_1)T_{p1}E_{pl}^*(t_2)R_{p2}^*$ equals ${E_{sl}(t_1)T_{s1}E_{sl}^*(t_2)R_{s2}^*}$,  there will be no interference pattern observed, even if there are interference pattern for parallel or perpendicular components individually.

For single-mode continuous-wave laser, the amplitude of electric field does not change with time. $|E_{sl}(t_1)|$ equals $|E_{sl}(t_2)|$ and $|E_{pl}(t_1)|$ equals $|E_{pl}(t_2)|$, which will be wrote as $|E_{sl}|$ and $|E_{pl}|$, respectively. Since only the spatial transverse part is studied in our experiments, $ \cos[\omega(t_1-t_2)] \cos[k_0(d_1-d_2)]$ is written as $B(\Delta z)$ for short, where $\Delta z$ is proportional to the light path difference between these two interfering light beams. The final first-oder interference pattern in the observation plane equals
\begin{eqnarray}\label{a6}
&&I_O\\
&=&|E_{pl}|^2[|T_{p1}|^2+|R_{p2}|^2+2\text{Re}(T_{p1}R_{p2}^*e^{i\varphi})B(\Delta z)\cos(2\Delta k x)]\nonumber\\
&&+|E_{sl}|^2[|T_{s1}|^2+|R_{s2}|^2-2\text{Re}(T_{s1}R_{s2}^*e^{i\varphi})B(\Delta z)\cos(2\Delta k x)],\nonumber
\end{eqnarray}
where $\varphi$ is the phase difference between electric field at $t_1$ and $t_2$. $|E_{pl}|^2$ and $|E_{sl}|^2$ are the intensities of laser light components with parallel and perpendicular polarizations, respectively. The terms within the square brackets are interference patterns depending on the transmission and reflection coefficients of  beam splitter. Once the incidental angle and beam splitters are determined, the visibility of the interference pattern can be calculated.

By changing the polarization of the incidental laser light, one can change the visibility of the interference pattern. It is different from the first-order interference pattern in the usual interferometer, such as Young\rq{}s double-slit interferometer, Michealson interferometer, M-Z interferometer, in which the first-order interference pattern is independent of the polarization of the incidental light.

\section*{Appendix B: Calculating the visibility of the first-order interference pattern}\label{appendix-b}
\def\theequation{$B-$\arabic{equation}}
\setcounter{equation}{0}
\def\thefigure{$B-$\arabic{figure}}
\setcounter{figure}{0}

In this section, we will show how to calculate the visibility of the interference pattern in Delta interferometer based on Eq. (\ref{I_of}). For simplicity, we assume $T_{p1}=R_{p2}$, $T_{s1}=R_{s2}$, $\varphi =0$, and $B(\Delta z)=1$, with which Eq. (\ref{I_of}) can be simplified as
\begin{eqnarray}\label{b1}
I_O&=&2|E_{pl}|^2|T_{p1}|^2[1+\cos(2\Delta k x)]\\
&&+2|E_{sl}|^2|T_{s1}|^2[1-\cos(2\Delta k x)].\nonumber
\end{eqnarray}
Equation (\ref{b1}) can be further simplified by assuming $T_{p1}=T_{s1}$, 
\begin{eqnarray}\label{b2}
I_O&\propto&|E_{pl}|^2+|E_{sl}|^2\\
&&+\cos(2\Delta k x)(|E_{pl}|^2-|E_{sl}|^2).\nonumber
\end{eqnarray}
When $|E_{pl}|^2-|E_{sl}|^2$ is larger than 0, the visibility of interference pattern can be expressed as
\begin{equation}\label{b3}
V_{EC}\equiv \frac{I_{Omax}-I_{Omin}}{I_{Omax}+I_{Omin}}=\frac{1-\frac{|E_{sl}|^2}{|E_{pl}|^2}}{1+\frac{|E_{sl}|^2}{|E_{pl}|^2}},
\end{equation}
where $I_{Omax}$ and $I_{Omin}$ are the maximum and minimum intensities of the interference pattern, respectively. With the help of Fig. \ref{b1}, Eq. (\ref{b3}) can be simplified as
\begin{equation}\label{b4}
V_{EC}=\frac{1-\tan^2 \phi}{1+\tan^2 \phi}.
\end{equation}

\begin{figure}[htb]
\centering
\includegraphics[width=45mm]{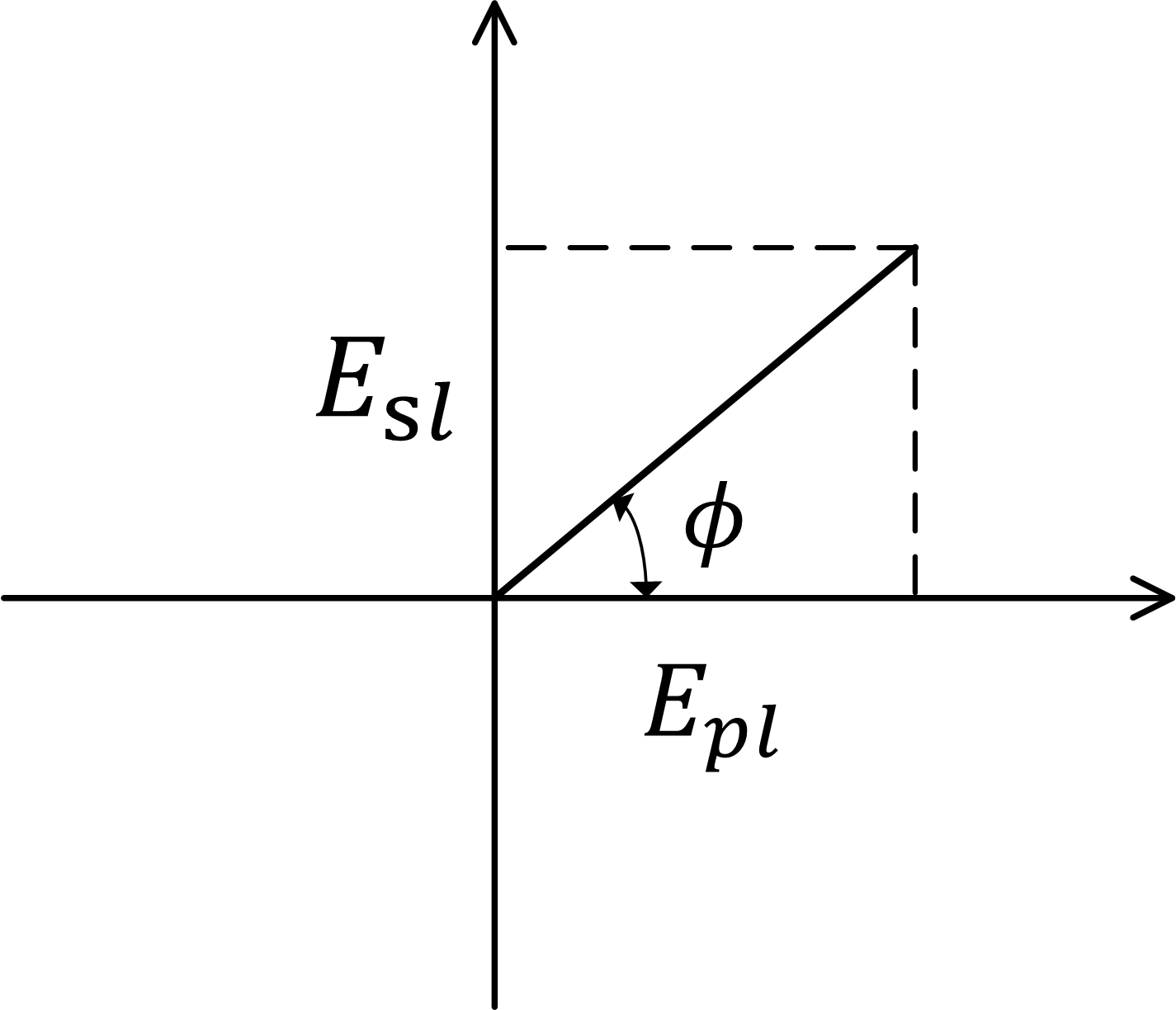}%
\caption{The relationship between polarization angle $\phi$ and the perpendicular and parallel electric field components. $E_{sl}$ and $E_{pl}$ are the perpendicular and parallel electric fields, respectively.}\label{b1} 
\end{figure}

With the same method above, it is easy to have the visibility of interference pattern when $|E_{pl}|^2-|E_{sl}|^2$ is less than 0,
\begin{equation}\label{b5}
V_{EC}=\frac{\tan^2 \phi-1}{\tan^2 \phi+1}.
\end{equation}
Hence the visibility of the interference pattern in Delta interferometer shown in Fig. \ref{4} equals
\begin{equation}\label{b6}
V_{EC}=\frac{|1-\tan^2 \phi|}{|1+\tan^2 \phi|},
\end{equation}
where polarization angle, $\phi$, is in the range of $(-\pi/2,\pi/2]$. For the polarization angle outside of $(-\pi/2,\pi/2]$, one can obtain the visibility based on the periodicity of triangle function.

Next, we will calculate the visibility of interference pattern in Delta interferometer for non-ideal condition, \textit{i.e.}, the visibility of interference pattern for $s$ or $p$ polarized light is not 100\%. Equation (\ref{b1}) should be re-written as
\begin{eqnarray}\label{b7}
I_O&=&2|E_{pl}|^2|T_{p1}|^2[1+\alpha \cos(2\Delta k x)]\\
&&+2|E_{sl}|^2|T_{s1}|^2[1-\alpha \cos(2\Delta k x)],\nonumber
\end{eqnarray}
where $\alpha$ is the visibility of $s$ or $p$ polarized light in Delta interferometer and is in the range of $[0,1]$. With the same method above, the visibility of the interference patter in Delta interferometer equals
\begin{equation}\label{b8}
V_{EC}=\frac{|1-\alpha \tan^2 \phi|}{|1+\alpha \tan^2 \phi|}.
\end{equation}
When $\alpha$ equals 1, Eq. (\ref{b8}) becomes identical to Eq. (\ref{b6}).


\begin{thebibliography}{99}


\bibitem{newton-optics} I. Newton, \textit{Opticks: Or a Treatise of the Reflexions, Refractions, Inflectxions and Colours of light} (Loudon, 1712).

\bibitem{young-1804} T. Young, Phil. Trans. R. Soc. Lond. \textbf{94}, 1 (1804).

\bibitem{aspect-1986} P. Grangier, G. Roger, and A. Aspect, Europhys. Lett. \textbf{1}, 173 (1986).

\bibitem{michelson} A. A. Michelson and E. W. Morley, Am. J. Sci. \textbf{34}, 333 (1887).

\bibitem{einstein} A. Einstein, Annal. Phys.  \textbf{322}, 891 (1905).

\bibitem{interferometer-book} P. Hariharan, \textit{Optical Interferometry (2nd ed.)} (Academic Press, CA, 2003).

\bibitem{EM-book1} J. D. Jackson, \textit{Classical Electrodynamics (3rd ed.) Chap. 7.3} (John Wiley \& Sons, Inc. MA, 1999).

\bibitem{EM-Yang} R. G. Yang and Y. L. Liu, \textit{Electric Field and Electrical Magnetic Wave (3rd ed.) (in Chinese) Chap. 8.8} (Advanced Education Press, Beijing, 2019).

\bibitem{pm-ao}H. Takasaki and Y. Yoshino, Appl. Opt. \textbf{8}, 2344 (1969).

\bibitem{pm-book} M. Fran\c{c}on and S. Mallic, \textit{Polarization interferometers: applications in microscopy and macroscopy} (Wiley-Interscience, 1971).

\bibitem{pm-rev} E. W. Ciurczak, Spectroscopy \textbf{20}, 68 (2005).

\bibitem{pm-cr} J. Escorihuela, M. \'{A}. Gonz\'{a}lez-Mart\'{i}nez, J. L. L\'{o}pez-Paz, R. Puchades, \'{A}. Maquieira, and D. Gimenez-Romero, Chem. Rev. \textbf{115}, 265 (2015).

\bibitem{pm-2000} H. Y. Stoyanov, Opt \& Laser Tech. \textbf{32}, 147 (2000).

\bibitem{pm-1987} P. Grangire, R. E. Slusher, B. Yurke, and A. LaPorta, Phys. Rev. Lett. \textbf{59}, 2153 (1987).

\bibitem{born-book} M. Born and E. Wolf, \textit{Principle of Optics (7th ed.)} (Cambridge University Press, Cambridge, 1999).
 
\bibitem{luo-2021} S. Luo, Y. Zhou, H. B. Zheng, W. T. Xu, J. B. Liu, H. Chen, Y. C. He, S. H. Zhang, F. L. Li, and Z. Xu, Opt. Express \textbf{29}, 30094 (2021).

\bibitem{handbook} M. Bass and V. N. Mahajan (ed.), \textit{Handbook of Optics (Vol. I): Geometrical and Physical Optics, Polarizd light, Components and Instruments (3rd ed.) Chap. 12.3} (The McGraw Hill Companies, Inc., NY, 2010). 



\end{thebibliography}
\end{document}